\documentclass[12pt]{article}
\usepackage{graphicx,rotating,color,hyperref,slashed}

\makeatletter
\newcounter{alphaequation}[equation]
\def\thealphaequation{\theequation\hbox to
0.6em{\hfil\alph{alphaequation}\hfil}}
\def\eqnsystem#1{
\def\@eqnnum{{\rm (\thealphaequation)}}
\def\@@eqncr{\let\@tempa\relax \ifcase\@eqcnt \def\@tempa{& & &} \or
\def\@tempa{& &}\or \def\@tempa{&}\fi\@tempa
\if@eqnsw\@eqnnum\refstepcounter{alphaequation}\fi
\global\@eqnswtrue\global\@eqcnt=0\cr}
\refstepcounter{equation} \let\@currentlabel\theequation \def\@tempb{#1}
\ifx\@tempb\empty\else\label{#1}\fi
\refstepcounter{alphaequation}
\let\@currentlabel\thealphaequation
\global\@eqnswtrue\global\@eqcnt=0 \tabskip\@centering\let\\=\@eqncr
$$\halign to \displaywidth\bgroup \@eqnsel\hskip\@centering
$\displaystyle\tabskip\z@{##}$&\global\@eqcnt\@ne
\hskip2\arraycolsep\hfil${##}$\hfil& \global\@eqcnt\tw@\hskip2\arraycolsep
$\displaystyle\tabskip\z@{##}$\hfil
\tabskip\@centering&\llap{##}\tabskip\z@\cr}
\def\endeqnsystem{\@@eqncr\egroup$$\global\@ignoretrue} \makeatother

\font\tenrsfs=rsfs10 at 12pt
\font\sevenrsfs=rsfs7
\font\fiversfs=rsfs5
\newfam\rsfsfam
\textfont\rsfsfam=\tenrsfs
\scriptfont\rsfsfam=\sevenrsfs
\scriptscriptfont\rsfsfam=\fiversfs
\def\mathscr#1{{\fam\rsfsfam\relax#1}}
\def\Lag{\mathscr{L}}
\makeatletter
\def\art{\@ifnextchar[{\eart}{\oart}}
\def\eart[#1]#2#3#4#5#6{{\rm #2}, {#3 \rm#4} {\rm (#6) #5} ({arXiv:#1})}
\def\hepart[#1]#2{{\rm #2, arXiv:#1}}
\newcommand{\oart}[5]{{\rm #1}, {#2  #3} {\rm (#5) #4}}

\newcommand{\be}{\begin{equation}}
\newcommand{\ee}{\end{equation}}

\def\bsg{\ifmmode B\to X_s\gamma\else $B\to X_s\gamma$\fi}
\def\bsll{\ifmmode B\to X_s\ell^+\ell^-\else $B\to X_s\ell^+\ell^-$\fi}
\def\shat{\ifmmode \hat{s}\else $\hat{s}$\fi}

\newcommand\m{\mu}

\newcommand\n{\nu}

\newcommand\s{\sigma}
\def\d{\partial}

\newcounter{mysubequation}[equation]

\newcommand{\TeV}{\,\mathrm{TeV}}
\newcommand{\GeV}{\,\mathrm{GeV}}
\newcommand{\MeV}{\,\mathrm{MeV}}

\def\beq{\begin{equation}}
\def\eeq{\end{equation}}
\def\bea{\begin{eqnarray}}
\def\eea{\end{eqnarray}}

\newcommand{\newc}{\newcommand}
\def\circa#1{\,\raise.3ex\hbox{$#1$\kern-.75em\lower1ex\hbox{$\sim$}}\,}

\newc{\gsim}{\lower.7ex\hbox{$\;\stackrel{\textstyle>}{\sim}\;$}}
\newc{\lsim}{\lower.7ex\hbox{$\;\stackrel{\textstyle<}{\sim}\;$}}

\def\eq#1{eq.~(\ref{eq:#1})}
\def\fig#1{fig.~\ref{fig:#1}}

\def\beq{\begin{equation}}
\def\eeq{\end{equation}}
\def\bea{\begin{eqnarray}}
\def\eea{\end{eqnarray}}

\begin{document}

\baselineskip=18pt

\setcounter{footnote}{0}
\setcounter{figure}{0}
\setcounter{table}{0}

\begin{titlepage}
\begin{flushright}
CERN-PH-TH/2011--234\\
\end{flushright}
\vspace{.3in}

\vspace{.5cm}

\begin{center}

{\Large\sc{\bf Interpreting OPERA results on superluminal neutrino}}

\vspace*{9mm}
\renewcommand{\thefootnote}{\arabic{footnote}}

\mbox{ \bf Gian F. Giudice$^{a}$, Sergey Sibiryakov$^{b}$,  Alessandro Strumia$^{c,d}$}

\vspace*{0.9cm}

$^{a}${\it  CERN, Theory Division,  CH--1211 Geneva 23, Switzerland}\\[3mm]
$^{b}${\it Institute for Nuclear Research of the Russian Academy of Sciences,\\
60th October Anniversary Prospect, 7a, 117312 Moscow, Russia}\\[3mm]
$^{c}${\it Dipartimento di Fisica dell'Universit{\`a} di Pisa and INFN, Italia}\\[3mm]
$^{d}${\it National Institute of Chemical Physics and Biophysics, Ravala 10, Tallinn, Estonia}\\[3mm]

\end{center}

\vspace{1cm}

\begin{abstract}
\medskip
\noindent\large
OPERA has claimed the discovery of superluminal propagation of neutrinos. We analyze the consistency of this claim with previous tests of special relativity. We find that reconciling the OPERA measurement with information from SN1987a and from neutrino oscillations requires stringent conditions. The superluminal limit velocity of neutrinos must be nearly flavor independent, must decrease steeply in the low-energy domain, and its energy dependence must depart from a simple power law. We construct illustrative models that satisfy these conditions, by introducing Lorentz violation in a sector with light sterile neutrinos. We point out that, quite generically, electroweak quantum corrections transfer the information of superluminal neutrino properties into Lorentz violations in the electron and muon sector, in apparent conflict with experimental data. 
\end{abstract}

\bigskip
\bigskip

\end{titlepage}

%%%%%%%%%%%%%%%%%%%%%%%%%%%%%%%%%%%%%%%%%%%%%%%%%%%
\section{Introduction}

The OPERA collaboration has recently announced the measurement of the velocity of neutrinos, as they travel from CERN to the Gran Sasso Laboratory (LNGS) covering a distance of about 730 km~\cite{opera}. The CERN Neutrino beam to Gran Sasso (CNGS) consists of $\nu_\mu$, with small contaminations of $\bar{\nu}_\mu$ (2.1\%) and of $\nu_e$ or $\bar{\nu}_e$ (together less than 1\%). The average neutrino energy is 17 GeV. OPERA has measured the neutrino velocity by taking the ratio between very accurate determinations of distance and time of flight. The distance is defined as the space separation
between the focal point of the graphite target (where the proton beam
extracted from the CERN SPS collides producing secondary charged
mesons that eventually decay into neutrinos) and the origin of the
OPERA detector reference frame. High-accuracy GPS measurements and
optical triangulations led to a determination of the distance with an
uncertainty of 20 cm (monitoring also Earth movements at the level of
centimeters). The real advance of OPERA with respect to previous
experiments lies in the time tagging system. An upgraded GPS-based
timing system at CERN and LNGS allows for time tagging with
uncertainties at the level of less than 10 nanoseconds. The neutrino
time of flight is then determined by a statistical comparison between
the distribution of the neutrino interaction time and the proton
probability density function computed from the known time structure of
the proton beam. The large data sample\footnote{In this paper we use
  the experimental results based on the original data sample
  reported in the version 1 of \cite{opera}. To remove a possible
  source of uncertainty, the OPERA collaboration has repeated the
  analysis with a reduced data sample of 15223 events. The final
  results presented in the version 2 of \cite{opera} are consistent
  with those quoted below within one standard deviation.} 
of 16111 neutrino events,
recorded in a 3-year period, brought the statistical error in the
analysis at the same level of the estimated systematic error. 

With this procedure, OPERA found the surprising result that neutrinos arrive earlier than expected from luminal speed by a time interval~\cite{opera}
\beq
\delta t = \left( 60.7 \pm 6.9_{\rm stat} \pm 7.4_{\rm syst} \right) {\rm ns}.
\eeq
This translates into a superluminal propagation velocity for neutrinos by a relative amount
\beq
\delta c_\nu =  \left( 2.48 \pm 0.28_{\rm stat} \pm 0.30_{\rm syst} \right) \times 10^{-5}~~~~~~{\rm (OPERA)},\label{eq:opr}
\eeq   
where $\delta c_\nu \equiv (v_\nu -c)/c$.
The same measurement was previously performed by MINOS (with 735 km baseline and a broad neutrino energy spectrum peaked around 3 GeV). Although not statistically significant, the MINOS result has a central value in the same ballpark of the recent OPERA determination~\cite{minos}
\beq
\delta c_\nu =  \left( 5.1 \pm 2.9 \right) \times 10^{-5}~~~~~~{\rm (MINOS)}.
\eeq   
Earlier short-baseline experiments have set upper limits on $|\delta
c_\nu |$ at the level of about  $4\times 10^{-5}$ in the energy range
between 30 and 200 GeV~\cite{old}.
 
The technical subtleties in the measurement of the neutrino time of
flight and the importance of this unexpected result  
call for further experimental scrutiny. Other experiments, such as
MINOS and T2K, are in a position to repeat the OPERA measurement and
this can give us decisive information. It is particularly interesting
that T2K has a baseline of only 295 km. Finding an early neutrino
arrival with a time interval of less than half of what measured by
OPERA would rule out a large class of unknown systematics that may
affect the present measurement.  

Waiting for further developments on the experimental side, theorists
can contribute to the debate by analyzing two questions. The first
issue is whether the OPERA measurement is already in conflict with
other tests of special relativity. The second question is whether we
can envisage simple deformations of the Standard Model that can be
reconciled with the measurement, possibly suggesting new ways of
testing experimentally these scenarios. In this paper, we  address
such questions.  

If the OPERA results are confirmed they will call for a complete reconsideration of the basic principles of particle physics. In a Lorentz-invariant theory, propagation of neutrinos faster than light would imply violation of causality ({\it i.e.}\ the CN$\,$GS beam becomes GS$\,$CN when seen from a sufficiently boosted reference frame). It is extremely hard to envisage a consistent theory having this property. An arguably less radical option is to admit the existence of a preferred reference frame that unambiguously defines causal relations but, by its very existence, breaks Lorentz invariance. 
In sect.~\ref{sum} we summarize and analyze the existing constraints on Lorentz violation relevant for our considerations. In sect.~\ref{fit} we fit the OPERA data and other experimental constraints with energy-dependent neutrino limit velocities.
In sect.~\ref{mod} we present possible models and their problems. Finally our conclusions are contained in
sect.~\ref{concl}.

\section{Constraints on Lorentz violation}\label{sum}

To assess the compatibility of the OPERA result on $\delta c_\nu$ with previous data we should consider three sets of constraints on Lorentz violation\footnote{For a recent review on
experimental constraints on Lorentz violation see ref.~\cite{Koseles}.}:
 {\it (i)} bounds from SN1987a; {\it (ii)} bounds on flavor-dependent  $\delta c_\nu$ from neutrino oscillations; {\it (iii)} bounds on $\delta c_\ell$, the corresponding limit velocity for charged leptons.

\subsection{Supernova SN1987a}
The detection of neutrinos emitted from SN1987a gave us plenty of information not only on the process of supernova explosion, but also on neutrino properties. The Irvine-Michigan-Brookhaven (IMB)~\cite{imb}, Baksan~\cite{baks}, and Kamiokande II~\cite{kam} experiments collected $8+5+11$ neutrino events
(presumably mainly $\bar\nu_e$)
with energies between $7.5$ and $39\MeV$ within 12.4 seconds. The time coincidence of these events can be used to constrain the difference in velocity of neutrinos with various energies. Indeed the time delay in the arrival of neutrinos with energy $E_1$ and $E_2$, respectively, is equal to
\beq
\delta t = \frac{d_{\rm SN}}{c}\left[\delta c_\nu (E_2) -\delta c_\nu (E_1) \right] ,
\eeq
where $d_{\rm SN}= 51$~kpc is the distance of SN1987a.
Rescaling a statistical analysis for the case of quadratic energy dependence, $\delta c_\nu \propto E^2$~\cite{Ellis}, we obtain the bound
\beq 
\left| \delta c_\nu (30\MeV) -\delta c_\nu (10\MeV) \right|
 < 5 \times 10^{-13}\hbox{ at 95\% CL}.\label{spr1}
 \eeq
However, a larger uncertainty could result from taking
into account that the average neutrino energy changes with time
during the detection interval of about 10 seconds. 

A direct constraint on $\delta c_\nu $, rather than on the difference between  $\delta c_\nu $ at different energies, can be derived from the observation that the optical signal from SN1987a came about 4 hours after the neutrino detection, in good agreement with supernova models. This implies 
\beq 
\left| \delta c_\nu (15\MeV)\right|   \label{spr2}
 \circa{<}10^{-9}\eeq
 This limit should be understood with an order-one uncertainty, which
 takes into account that the precise time delay between light and neutrinos is not known.
 
\begin{figure}
$$\includegraphics[width=0.70\textwidth]{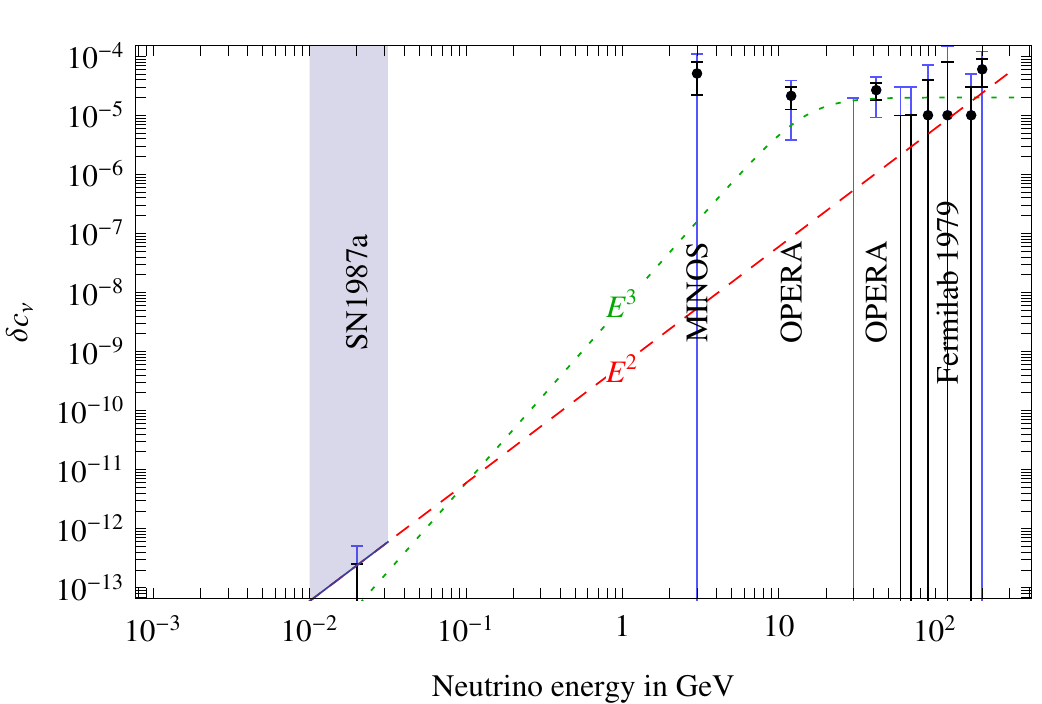}$$ 
$$\includegraphics[width=0.70\textwidth]{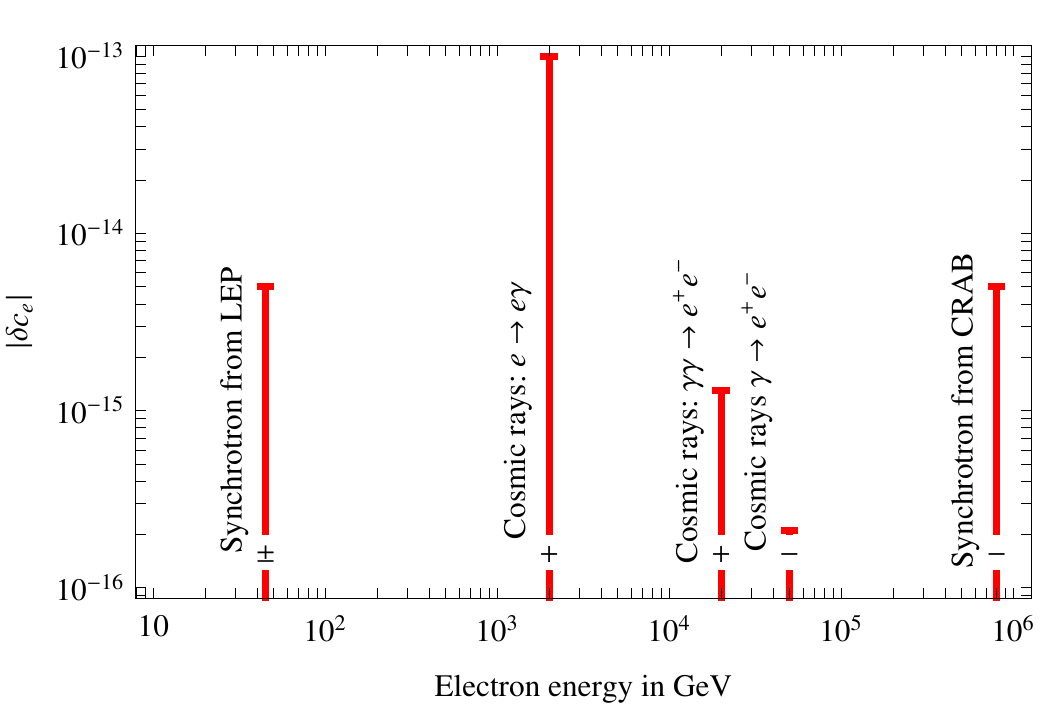}$$
\caption{\em Top: MINOS, OPERA and {\sc FermiLab1979} data on $\delta c_\nu$, together with the upper limit from SN1987a (shaded region). Error bars are at $1\sigma$ (black) and $2\sigma$ (blue). Overlaid are best fits for
 $\delta c_\nu \propto E_\nu^2$ (dashed red line) and $\delta c_\nu \propto E_\nu^3/(E_\nu^3+ E_*^3)$ (dotted green line).
Bottom: Upper limits on $|\delta c_e|$.
The bounds marked with ``+'' (``$-$'') apply to the case of positive (negative) $\delta c_e$. See text for an explanation of the different bounds.
\label{fig:data}}
\end{figure}

 In summary, limits on $\delta c_\nu$ from SN1987a are many orders of magnitude more stringent than the positive signal of OPERA. 
One way to evade the SN1987a bound would be to assume that $\delta c_\nu$ is sizable and positive for neutrinos, but vanishes for antineutrinos. Since the detectors that measured the supernova flux were mainly sensitive to antineutrinos, the SN1987a observation would not contradict the large $\delta c_\nu$ claimed by OPERA. However, we are not aware of a concrete setup that realizes this possibility. A more viable solution exploits the fact that
the SN1987a limits are not directly applicable to the OPERA result, since they refer to a different neutrino energy domain,
 as summarized in \fig{data}a.

\subsection{Neutrino oscillations}
\label{sec22}
The effective Hamiltonian describing neutrino oscillations can be written as
\beq 
H_{\rm eff} = \left( 1+ \delta c_\nu \right)  E + \frac{m^2}{2E} . 
\eeq
Here $m^2$ is the usual $3\times 3$ neutrino mass matrix and $\delta c_\nu $ describes the Lorentz violating effect. In general  $\delta c_\nu $ too could be a matrix in flavor space. The observation that neutrino oscillations follow the expected pattern with oscillation phases proportional to $L/E$, where $L$ is the distance travelled by neutrinos, gives strong bounds on flavor-dependent Lorentz violations. 

Off-diagonal elements of the matrix $\delta c_{\nu_i \nu_j}$ are constrained~\cite{ColemanGlashow, boo} and must satisfy the bound
\beq
 \left| \delta c_{\nu_i \nu_j}\right| < \frac{1}{EL} \approx2\times 10^{-19} \left( \frac{\GeV}{E}\right) \left( \frac{{\rm km}}{L}\right) .
 \eeq 
Differences between diagonal elements in the neutrino mass eigenstate basis
are similarly constrained, because
oscillations involving  active neutrinos have been observed and the oscillation dip has
been clearly seen in KamLAND ($\bar\nu_e\to \bar\nu_{e}$ disappearance at
$L\sim 100\,{\rm km}$, $E\sim 10\MeV$ due to solar $\nu_e\to \nu_{\mu,\tau}$ oscillations~\cite{review}) as
well as in K2K, MINOS and T2K~\cite{T2K} energy spectra
($\nu_\mu$ disappearance due to $\nu_\mu\to \nu_\tau$ atmospheric oscillations 
at $L\sim 300\,{\rm km}$ and $E\sim 3\GeV$).  
Too large neutrino velocity differences would split the wave-packets thereby destroying coherence and erasing the observed oscillation dips. Then we obtain the bound
\beq 
 \left| \delta c_{\nu_i \nu_i}-\delta c_{\nu_j \nu_j}\right|
\circa{<}10^{-(19\div 21)}\ ,\eeq
where the exact value of the exponent depends on the indices $i,j$.

In summary, flavor-dependent Lorentz violations in the neutrino sector are so strongly constrained that we can hope to explain the OPERA measurement only with a universal limit velocity for all neutrinos.    
The same conclusion applies to other Lorentz-breaking operators that affect the neutrino velocity. Therefore, it seems reasonable to assume that $\delta c_\nu$ is proportional to the identity matrix in flavor space.

\begin{figure}
$$\includegraphics[width=0.43\textwidth]{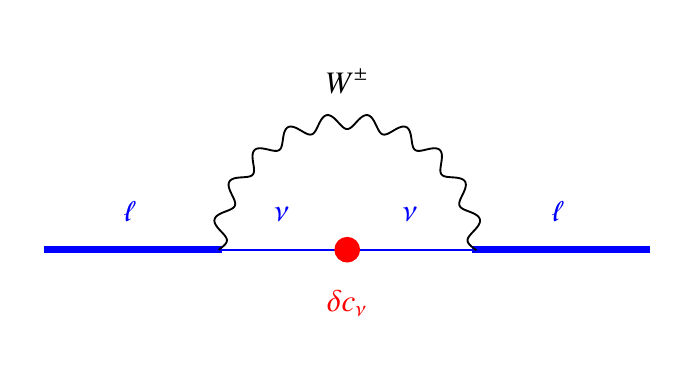}$$\vspace{-1cm}
\caption{\em The Feynman diagram exhibiting how Lorentz violation in neutrinos, represented by the red blob, is transmitted to charged leptons $\ell$
by quantum electroweak corrections.
\label{fig:Feyn}}
\end{figure}

\subsection{Charged leptons}
Left-handed neutrinos are part of electroweak doublets. Thus gauge interactions can relate Lorentz-violating effects in neutrinos and in charged leptons. This can be a serious concern for any attempt to explain the OPERA measurement in a consistent theoretical framework, because the present limits on $\delta c_e $ for electrons are much more stringent than for neutrinos.
In \fig{data}b we summarize the bounds on $\delta c_e $, relating them to the relevant electron energy scales. The labels in \fig{data}b correspond to the following observations.

\begin{itemize}
\item ``Synchrotron from LEP" is the limit $|\delta c_e|< 5\times 10^{-15}$ derived from agreement between observation and the theoretical expectation of electron synchrotron radiation as measured at LEP~\cite{bou1}. 

\item ``Cosmic rays: $e\to e \gamma$" refers to the observation that electron vacuum $\check{\rm C}$erenkov radiation (the decay process $e\to e \gamma$) becomes kinematically allowed for $E_e >m_e/\sqrt{\delta c_e}$. Since cosmic ray electrons have been detected up to 2~TeV, one deduces 
$\delta c_e <  10^{-13}$~\cite{bou2}. 

\item ``Cosmic rays: $\gamma \gamma\to e^+e^-$" refers to the process in which high-energy photons are absorbed by CMB photons and annihilate into an electron-positron pair. The process becomes kinematically possible for $E_{\rm CMB} > m_e^2/E_\gamma +\delta c_e E_\gamma/2$. Since it has been observed to occur for photons of about $E_\gamma =20$~TeV, this implies $\delta c_e <2m_e^2/E_\gamma^2 \sim 10^{-15}$~\cite{bou3}. 

\item ``Cosmic rays: $\gamma \to e^+e^-$" refers to the process in which a photon decays into $e^+e^-$,
which becomes kinematically allowed at energies $E_\gamma > m_e\sqrt{-2\delta c_e}$.  Since photons have been observed 
up to 50 TeV one deduces $-\delta c_e < 2\times 10^{-16}$~\cite{bou2}.
The analogous bound for the muon is $-\delta c_\mu < 10^{-11}$~\cite{Altmu}.

\item ``Synchrotron from CRAB" refers to the limit $-\delta c_e< 5\times 10^{-15}$ deduced from astrophysical observations of inverse Compton scattering and synchrotron radiation~\cite{Alts}.
\end{itemize}

At first sight it may seem that the strong limits on $\delta c_e$ could be evaded by introducing a Lorentz violating interaction with appropriate Higgs insertions able to select only neutrino fields. One such example is the space rotationally invariant operator $(H^\dagger L^\dagger)\partial_0 (LH)$, where $L$ is the lepton doublet and SU(2) indices are saturated between the fields in parenthesis. Below the weak scale, the Higgs vacuum expectation value leads to an effective operator involving neutrinos but not charged leptons. However, quantum corrections due to weak interactions generate a non-vanishing  $\delta c_e$ proportional to the original $\delta c_\nu$. This can be seen in the previous example, since the operator $(H^\dagger L^\dagger)\partial_0 (LH)$ has an anomalous dimension mixing with the operator $(H^\dagger H)L^\dagger\partial_0 L$, which contributes to $\delta c_e$.

The effect is quite general and comes from a one-loop diagram with internal exchange of a $W$ boson and a neutrino with Lorentz-violating insertion (see fig.~\ref{fig:Feyn}), giving
\beq 
 \delta c_e(p) \approx g^2 \int \frac{d^4k}{(2\pi)^4} \frac{\delta c_\nu(k)}{k^2[(k+p)^2-M_W^2]}.
 \label{loop}
\eeq
Barring cancellations, the most optimistic situation occurs when $\delta c_\nu$ is as large as the value measured by OPERA in a small energy range around $E_{\rm OPERA}\sim 20\GeV$ and vanishes elsewhere. (In sect.~\ref{mod} we will construct examples of models that fulfill these conditions). In this way, from eq.~(\ref{loop}) we can set a generic lower limit on the value of $\delta c_e$ based on the OPERA measurement:
\beq
 \delta c_e
\circa{>} \left( \frac{E_{\rm OPERA}}{4\pi v}\right)^2 \delta c_\nu (E_{\rm OPERA})\sim 10^{-9}.\label{eq:estimate}
\eeq
This result is clearly in conflict with the limits summarized in \fig{data}b.

To justify eqs.~(\ref{loop}) and~(\ref{eq:estimate})
we consider a simple toy model, with a Yukawa interaction between an SU(2) lepton doublet and a sterile neutrino $N$
with Lorentz-violating speed $c(1 + \delta c_N)$. This simplified model illustrates well the dynamics of how $\delta c_e$ is generated, reproducing all the essential features of the more realistic examples later discussed in sect.~\ref{mod}. The Lagrangian of the toy model is
\beq  \Lag = \Lag_{\rm SM} + \bar N i \tilde{\slashed{\partial}}  N+ \lambda HLN  +
\frac{M}{2} N^2
\label{toytoy}
\eeq
where $\tilde{k}_\mu \equiv k_\mu - \delta c_N u_\mu (k\cdot u)$ for any quadri-vector $k$
and $u_\mu = (1,0,0,0)$.
At tree level one finds $\delta c_\nu(k) = (\lambda v/M)^2 \delta c_N$ up to $k<M/\sqrt{\delta c_N}$,
while $\delta c_\nu$ is suppressed by $1/k^2$ at large $k$.

At high energies, much above the $W$ mass,
the Yukawa interaction 
generates at the one-loop level a correction to the limit velocity of the left-handed
electron (and neutrino) in the $L$ doublet:
\beq \delta c_L =\frac{ \lambda^2\delta c_N}{6(4\pi)^2}\bigg[\frac{1}{\epsilon}+\ln \frac{\bar\mu^2}{\Lambda^2}+{\cal O}(1)\bigg]
\label{high-E}
\eeq
where we used  regularization in $d=4-2\epsilon$ dimensions.
In the $\overline{\rm MS}$ renormalization scheme the $1/\epsilon$ pole is reabsorbed in the definition of $\delta c_L$
and the RGE scale $\bar\mu$ is identified with the low energy scale, $\bar\mu\sim M_W$.
Here $\Lambda$ is the maximal scale at which the theory is valid, which 
conservatively could be as low as
$\Lambda \sim M_W$.

At low energies we can compute the effect by evaluating the contribution of $W$ boson exchange using the unitary gauge,
such that the only diagram is the one shown in \fig{Feyn}. 
Neglecting fermion masses ($p^2=0$), the correction to the electron self-energy $\Sigma_e(p) = \slashed{p}$ is
\beq  \delta\Sigma_e(p)= 
\frac{g^2(\lambda v)^2}{2} \int \frac{d^4k}{(2\pi)^4} \frac{\gamma_\mu  \slashed{k}\tilde{\slashed{k}}\slashed{k}\gamma_\nu P_L}{k^4\tilde k^2} 
\frac{[ -g_{\mu\nu} +(k+p)_\mu (k+p)_\nu/M_W^2]}{(k+p)^2-M_W^2}.
\label{loopint}
\eeq
%-\frac{1}{3}
%g^2 \gamma_0 p_0 \int \frac{d^4k}{(2\pi)^4}\frac{1+{\cal O}(k^2/M_W^2)}{[k^2-M_W^2]^2}.\eeq
Here we are interested in the Lorentz-violating term inducing a non-vanishing $\delta c_e$, which is given by the part of $\delta\Sigma_e$ proportional to $\slashed{u}$.
At leading order in $\delta c_N$, the neutrino propagator acquires a Lorentz-violating term proportional to
\beq \frac{1}{\slashed{k}} - \delta c_N\bigg[ \slashed{u}\frac{k\cdot u}{k^2} - 2 \slashed{k} \frac{(k\cdot u)^2}{k^4}\bigg]\label{eq:ins}\eeq
and we find a modification of the limit velocity of left-handed electrons,
\beq \delta c_{e}(p)= \frac{g^2}{2(4\pi)^2} \frac{(\lambda v)^2\delta c_N}{M_W^2}\bigg\{1+\frac{1}{6}(\frac{1}{\epsilon} +
 \ln\frac{\bar\mu^2}{M_W^2}-\frac{1}{6})\bigg\}
\bigg[1 + \delta c_N {\cal O} (\frac{p\cdot u}{M_W})^2\bigg] .
\label{quindici}
\eeq
The first term inside the curly brackets comes from transverse $W$ polarizations and the
second (divergent) term comes from the longitudinal $W$ polarization
and reproduces the effect in eq.~(\ref{high-E}). 
 The lowest-order term agrees with eq.~(\ref{eq:estimate}) after taking into account that one needs $M$ in the same energy range as $E_{\rm OPERA}$,
while higher-order terms are model-dependent.
Indeed, in this model we obtain
\beq  \delta c_e(p \ll \frac{M_W}{\sqrt{\delta c_N}})= \delta c_e(p)|_{\overline{\rm MS}} + \frac{35}{36}\frac{M^2}{(4\pi v)^2}\delta c_\nu .
\label{nuova}
\eeq
This result holds as long as $M \ll 4\pi v$ (which is the regime relevant for the models considered in sect.~\ref{mod}) because we have used the tree-level relation between $\delta c_N$ and $\delta c_\nu$.  For $M \gg 4\pi v$, the one-loop correction becomes dominant and eq.~(\ref{nuova}) should be modified.

In the limit of large electron energy ($p^2=0$ with $p\cdot u\gg M_W/\sqrt{\delta c_N}$) the effect can be more easily extracted from the  unbroken SU(2) limit. Since the loop function of the integral in eq.~(\ref{loopint}) depends on the single variable $p\cdot u /M_W$, the high-energy limit is obtained by setting $M_W =0$, which corresponds to the calculation in eq.~(\ref{high-E}).
In this regime, for dimensional reasons, $\delta c_e$ remains roughly constant, with a logarithmic dependence on $p/M_W$, such that the effect cannot be fine-tuned away.

\bigskip

One may believe that the correlation between $\delta c_\nu$ and $\delta c_e$ is specific to the interaction in eq.~(\ref{toytoy}). To show that this is not the case, let us consider a different way of breaking Lorentz invariance, through a $\Delta L = 2$ term
$\frac{1}{2}\kappa   [Ni(\partial\cdot u) N + \hbox{h.c.}]$, which corresponds to an energy-dependent 
Majorana `mass-like' term $m_N = \kappa (k\cdot u)$.

The modification of the neutrino velocity is then given by $\delta c_\nu = (\lambda v/M)^2 \kappa^2$, 
where the square appears because we consider lepton-number conserving
neutrino-to-neutrino propagation.
For the same reason, the correction to the electron limit velocity is also proportional to $\kappa^2$ and
consequently again proportional to $\delta c_\nu$, as in \eq{estimate}.
Indeed, the only change in the loop contribution is that \eq{ins} becomes
\beq \frac{1}{\slashed{k}} - \kappa^2 \slashed{k} \frac{(k\cdot u)^2}{k^4}\label{eq:inspipi}\eeq
where the Lorentz-breaking term has the same form as the last term in \eq{ins}.
Given that both Lorentz-breaking terms  in \eq{ins} give non-vanishing contributions to $\delta c_e$
with comparable order one coefficients and that no other terms are possible, we 
conclude that the estimate of \eq{estimate} has general validity.

%Assuming the dispersion relation $E(1+f)  = p$ such that $\delta c  = f + d f/d\ln E$ we get, using the unitary gauge:
%\beq 
%f_e(p=0) = g^2 \int \frac{d^4k}{(2\pi)^4} \frac{k^2/M_W^2-2}{k^2[k^2-M_W^2]} f_\nu(k).
% \label{loop2}
%\eeq

The muon limit velocity $\delta c_\mu$ is less constrained than in the case of the electron~\cite{Altmu,g-2}. However the strong bounds on Lorentz violation for non-universal flavor interactions prevents us from restricting superluminal effects to the $\nu_\mu$--$\mu$ sector. 

In summary, the existing limits on $\delta c_e$ present a formidable obstacle for a consistent interpretation of the OPERA result. 

Moreover, the unavoidable non-universality due to the charged lepton masses can propagate at the quantum level an original universal $\delta c_\nu$ into flavor-dependent contributions. Given the strong bounds discussed in sect.~\ref{sec22}, this could also create serious difficulties in the constructions of viable theories.

\begin{figure}
$$\includegraphics[width=0.43\textwidth]{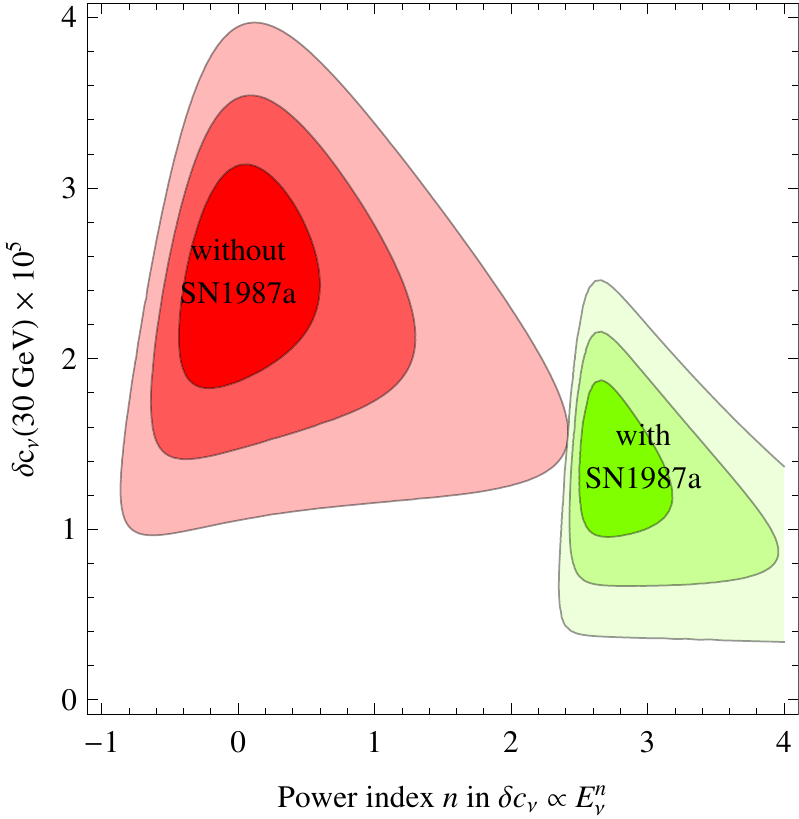}\qquad\includegraphics[width=0.45\textwidth]{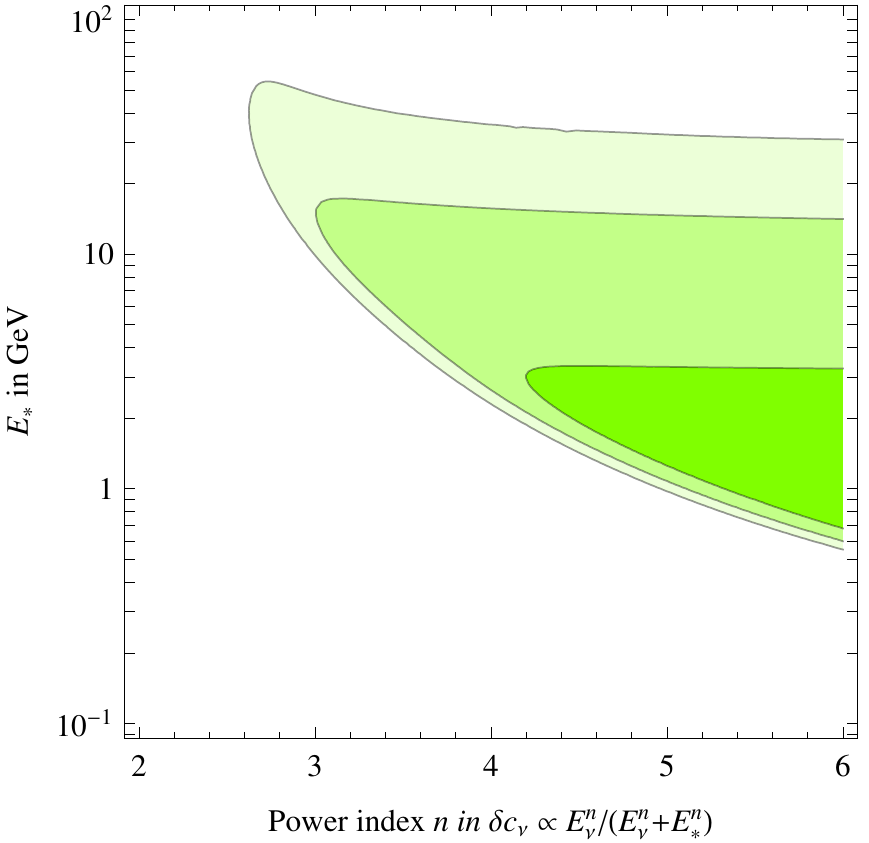}$$
\caption{\em 
Left: Fit of the OPERA and MINOS data assuming $\delta c_\nu \propto E_\nu^n$. The red contours do not include the SN1987a bound (and have $\chi^2_{\rm best} = 1$), while the green ones do (with $\chi^2_{\rm best} = 14$). The contours are at $1,2$ and $3\sigma$ (68\%, 95\%, 99.73\% CL for 2 dof). Right: Same fit assuming $\delta c_\nu \propto E_\nu^n/(E_\nu^n+E_*^n)$ and including the SN1987a bound.
\label{fig:fit}}
\end{figure}

%\section{Model-independent analysis}   
%
%In \fig{fit}a we fit the OPERA and MINOS data assuming $\delta c_\nu \propto E_\nu^n$.
%Data do not show any sign of an energy dependent effect, such that an
%energy-independent $\delta c_\nu$ is preferred.
%The red curves around $n\approx 0$ shows the corresponding best-fit region, altough 

\section{Model-independent analysis}   \label{fit}

The OPERA measurement, \eq{opr}, together with the bounds from SN1987a in eqs.~(\ref{spr1})--(\ref{spr2}) and those from neutrino oscillations, imply that $\delta c_\nu$ must be nearly flavor-universal and must show a steep energy dependence. It is not difficult to imagine a power law that gives $\delta c_\nu \propto E_\nu^n$, with $n$ positive. The problem however lies in the OPERA data themselves that show no evidence for an energy dependence in the available energy domain. Using only charged-current events inside the detector (a sample of 5489), the collaboration has presented an analysis in which the data are split into two bins of comparable statistics, containing events of energy larger or smaller than 20 GeV. The average energies of the two bins are 13.9 and 42.9~GeV, respectively. 
The time of neutrino early arrival for the two bins is measured to be~\cite{opera}
\bea
&&\delta t = \left( 53.1 \pm 18.8_{\rm stat}  \right) {\rm ns} ~~~{\rm for}~\langle E_\nu \rangle = 13.9\GeV \nonumber \\
&&\delta t = \left( 67.1 \pm 18.2_{\rm stat}  \right) {\rm ns} ~~~{\rm for}~\langle E_\nu \rangle = 42.9\GeV .
\label{datbin}
\eea
The difference between the two values is $(14.0 \pm 26.2)$~ns, indicating no significant energy dependence.

The situation is illustrated in \fig{data}a, where the OPERA and MINOS measurements are plotted together with the SN1987a constraint. While the large value of $\delta c_\nu$ from OPERA requires a power index bigger than 2, such a steep energy dependence is in conflict with the two-bin analysis. This conflict is quantified in \fig{fit}a, where we fit the data assuming a power law $\delta c_\nu \propto E_\nu^n$. In the global fit we take into account the energy dependence of the OPERA data by appropriately combining the results in eq.~(\ref{datbin}).

 The red region in \fig{fit}a shows the fit obtained including only OPERA and MINOS data:
the best fit has a very good $\chi^2\approx 1$ and a power-law index $n\approx 0$ is preferred.

The green region shows the global fit, where also the SN1987a data are included.
A power-law index $n\le 2$ is strongly disfavored.  The two regions are statistically incompatible, 
and indeed the global fit\footnote{Of course only the OPERA collaboration can perform a precise fit. However, although details could change, the main features of our analysis are robust.} including SN1987a has a high $\chi^2 \approx 14$.
Furthermore, the global fit is incompatible with the older measurements at higher energies~\cite{old},
shown in figure~\fig{data} collectively as ``{\sc FermiLab1979}'', because of the  growth of $\delta c_\nu$ at high energy.

In order to obtain a good global fit, one  must assume a distortion from a simple power law for $\delta c_\nu$. This situation is less conventional, since it requires the existence of some new physics threshold between the energies of SN1987a (about 10~MeV) and OPERA (about 20~GeV). The same conclusion has also been reached in ref.~\cite{newpap}.

In sect.~\ref{mod} we present examples of models that can produce a distortion of the power law, as suggested by data.
Here, we assume a simple change from power-law behavior at low energies to constant behavior at high energies, $\delta c_\nu \propto E_\nu^n /(E_\nu^n + E_{ *}^n)$. As shown in \fig{fit}b, a statistically acceptable fit is obtained for $n\ge 3$ with $E_{ *}\sim 10$~GeV. MINOS data give a 1-$\sigma$ preference towards smaller values of  $E_{ *}$ and steeper energy growth ($n\ge 4$). 
An example of such a fit that  well interpolates between data and constraints is shown in  \fig{data}a, where we assume
$n=3$ and $E_{ *}= 15$~GeV.

\section{Models}\label{mod}
We want to study now how to generate a $\delta c_\nu$ with an energy functional dependence different than a power law. The safest way is to introduce one or more sterile neutrinos $N_i$ with masses $M_i$ within the OPERA energy range and confine the Lorentz-violating interactions in this sector. 
The relevant Lagrangian is
 \beq \Lag = \Lag _{\rm SM} + \sum_i  \left[ i\bar N_i {\bar \sigma}^\mu \partial_\mu N_i
  +\left( \frac{M_i}{2} N_i^2 + \lambda_i H LN_i + \hbox{h.c.}\right)\right]+\Lag_{\rm Lorentz}.
 \label{lagrr}
 \eeq
 Here $\Lag_{\rm Lorentz}$ involves only the fields $N_i$ and violates Lorentz invariance.
The Yukawa couplings $\lambda_i$ must be identical for any flavor of the doublet $L$, but additional contributions to neutrino masses may be present in $\Lag _{\rm SM}$.
 As sterile neutrinos have no gauge interactions, 
Lorentz breaking will be transmitted to active neutrinos in the lepton doublets $L$
at tree level via the active/sterile mixing angles $\theta_i \simeq  \lambda_{i} v/M_i\ll1$
 (experimentally constrained to be $\theta_i <0.1$) and to  charged leptons at loop level.

In the low-energy limit ($E \ll M_i$), the light state mostly coincides with the active one 
\beq \nu_{\rm light} = \nu + \sum \theta_i N_i + {\cal O}(E/M_i). \eeq
In order to be compatible with the SN1987A bounds, 
$\Lag_{\rm Lorentz}$ must have zero projection over the light state at low energies.
Assuming, for example, energy-independent Lorentz breaking terms $\delta c_{ij}$ in the $N_i$--$N_j$ kinetic mixings, one has
\beq \delta c_\nu(E\ll M_i) = \sum_{i,j} \theta_i \theta_j \delta c_{ij} + {\cal O}(E^2/M_i^2) ,
\eeq
which vanishes for a sector of two sterile neutrinos where the only non-zero terms are $\theta_2$ and $\delta c_{12}$.
Assuming that only one of the two sterile neutrinos mixes with the active neutrino, and that
Lorentz-breaking is only present in the sterile neutrino mixing,
we automatically find that at low energy  
$\delta c_\nu (E) \propto E^n$ with  $n\ge 2$. The power suppression at low energy can account for the absence of anomalous effects in SN1987a.

\begin{figure}
$$\includegraphics[width=0.43\textwidth]{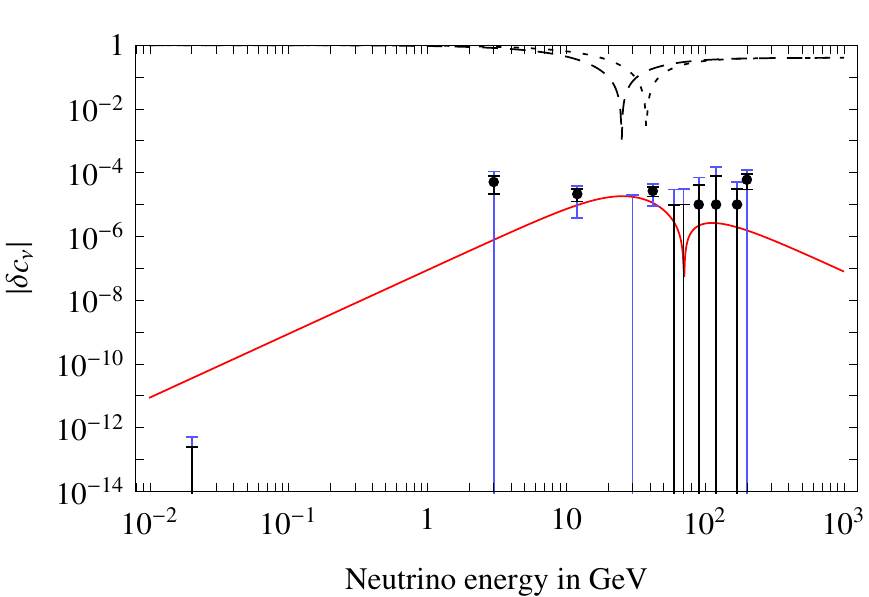}\qquad\includegraphics[width=0.45\textwidth]{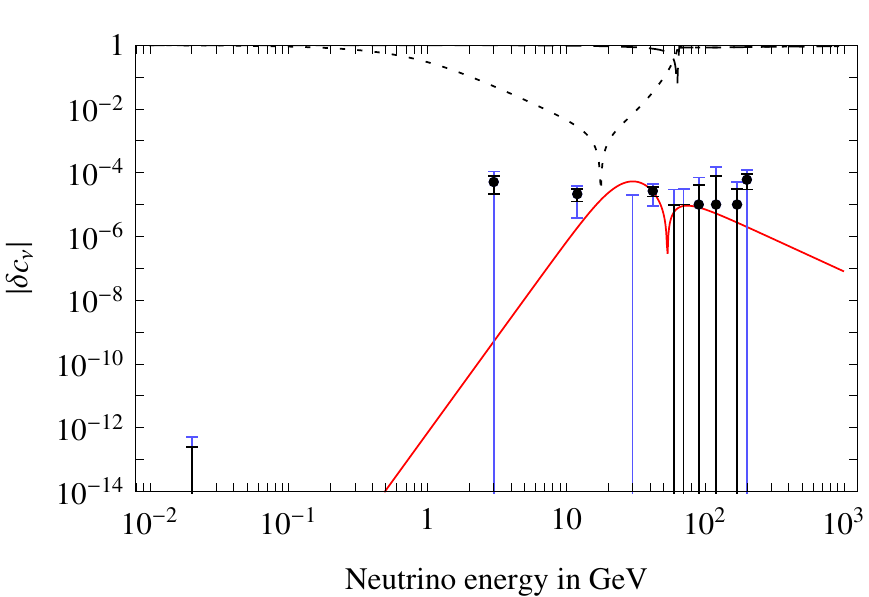}$$
\caption{\em 
$|\delta c_\nu| $ of the active neutrino (solid red curve) and of the heavy extra sterile neutrinos
that play no active role (dashed and dotted black curves).  The left (right) frame refers to the model 
of section~\ref{mod1} (\ref{mod2})
with the parameter choice described in the text.   The dips indicate changes of sign of $\delta c_\nu$, where $\delta c_\nu$ for the active neutrino is positive for energies in the OPERA range and below, and negative at higher energies.
\label{fig:mod}}
\end{figure}

\subsection{A model with 2 sterile neutrinos}
\label{mod1}

To protect the SM sector from violation of Lorentz
invariance we will adopt the framework of ref.~\cite{susy}
 based on
supersymmetry. In this context no Lorentz violating 
operators of dimension
less or equal to four are allowed for the superfields describing SM particles. However, a dimension-4 Lorentz
breaking kinetic mixing 
for gauge singlets (such as sterile neutrinos) is allowed. In
the superfield language this operator has the form 
\beq
\label{super}
\frac{i}{2}\kappa_{ij}u^\m\int d^2\theta\; N_{i}\d_\m N_{j}+\mathrm{h.c.}
\eeq
Here $N_{i}$ are sterile neutrino chiral superfields with the index $i$
labeling species.
The fixed vector $u^\m$ with the components
\beq
\label{umu}
u^\m=(1,0,0,0) 
\eeq
is introduced for convenience to describe Lorentz breaking in a
formally covariant way, where space rotational invariance is retained. Finally, the coupling constants $\kappa_{ij}$
form an antisymmetric matrix, because the symmetric part of the interaction is a total derivative. Thus, this interaction is absent in the case of a single field, and one off-diagonal term is the only possible interaction in the case of two fields. Here we consider the case of two sterile neutrinos.

The operator (\ref{super}) produces the kinetic mixing for the two
sterile neutrinos\footnote{We work with two-component spinors 
and follow the conventions of ref.~\cite{WB}. The signature of the metric is
$(-,+,+,+)$.},
\beq
\label{buf}
\Lag_{\rm Lorentz} =i\kappa \, u^\m N_1 \d_\m N_2-i\kappa^*\, u^\m \bar N_1\d_\m \bar N_2\;.
\eeq
Thus the full neutrino Lagrangian including mixings with the
left-handed neutrino $\n$ is
\beq
\label{fullL}
\begin{array}{rl}
\Lag=& i\bar\n\bar\s^\m \d_\m\n+i\bar N_1\bar\s^\m\d_\m N_1+i\bar N_2\bar\s^\m\d_\m N_2
+\bigg(\displaystyle \frac{M_1}{2}N_1^2+\frac{M_2}{2}N_2^2+m_1\n N_1+m_2 \n N_2+\mathrm{h.c.} \bigg)
\\
&+i\kappa \, u^\m N_1 \d_\m N_2-i\kappa^*\, u^\m \bar N_1\d_\m \bar N_2\;,
\end{array}
\eeq
where $m_{1,2}=\lambda_{1,2}v$ and we take $M_1<M_2$.

In the high-energy regime ($E\gg M_{1,2}$), the sterile and active neutrinos decouple and $\delta c_\nu$ is suppressed. At intermediate energies ($M_1\ll E\ll M_2$) we can integrate out the heaviest of the two sterile neutrinos, finding
\beq
\label{LhighE}
\begin{array}{rcl}
\Lag &=& i\bar\n\bar\s^\m \d_\m\n+
(m_1\n N_1+\mathrm{h.c.})+i\bar N_1\bar\s^\m\d_\m N_1
+\bigg(\displaystyle\frac{M_1}{2}N_1^2+\mathrm{h.c.}\bigg)\\
&&\displaystyle -\frac{m^2_2}{2M_2}\n^2+i\frac{\kappa m_2}{M_2} u^\m\n\d_\m N_1
+\mathrm{h.c.}\;,
\end{array}
\eeq
where we have kept only terms with up to one derivative.
There is a non-trivial kinetic mixing which, neglecting masses and solving
for the equations of motion, gives
\beq
\label{vnu}
\delta c_\nu =|\delta|^2\;, ~~~~~\delta\equiv \frac{\kappa m_2}{M_2}\;.
\eeq
The quadratic dependence of $\delta c_\nu$ on $\kappa$ arises because the operator of eq.~(\ref{buf}) violates lepton number,
while OPERA observes $\nu_\mu\to\nu_\mu$ propagation.

It is interesting to find that the sign of $\delta c_\nu$ is fixed and thus neutrinos are necessarily superluminal. This is a direct consequence of eq.~(\ref{buf}), which predicts sterile neutrinos faster than light, while the property of being superluminal is preserved in mixing. The value of $\delta c_\nu$ in eq.~(\ref{vnu}) is independent of energy, offering the possibility of explaining the flattening of the neutrino speed in the OPERA energy range. In practice, however, the intermediate energy regime ($M_1\ll E\ll M_2$) should be very small in order to avoid large mixing angles between sterile and active neutrinos.

In the low-energy regime ($E\ll M_1, M_2$) we can integrate out both sterile
neutrinos, getting the effective Lagrangian up to one derivative
\beq
\label{LlowE}
\Lag=i\bar\n\bar\s^\m \d_\m\n
+\bigg(\frac{\kappa m_1 m_2}{M_1 M_2}iu^\m\n\d_\m\n
-\frac{m_1^2}{2M_1}\n^2-\frac{m_2^2}{2M_2}\n^2+\mathrm{h.c.}\bigg)\;,
\eeq
We see that apart from the standard neutrino masses we obtain a Lorentz violating
kinetic term. However, for a single left-handed neutrino this term is just
a total derivative. We should then consider higher-derivative Lorentz-violating interactions. Taking for simplicity $m_1=0$, we find 
\beq
\label{LVquadr}
\Lag_{LV}=\frac{\kappa^2 m_2^2}{2M_1 M_2^2}(u^\m\d_\m\n)^2+\mathrm{h.c.}
\eeq
This leads to a quartic dispersion relation at low energy:
\beq
\label{disp}
E^2-p^2-\frac{|\kappa|^4 |m_2|^4}{M_1^2 M_2^4}E^4=0\;.
\eeq
Thus we obtain
\beq
\delta c_\nu =\frac{3|\kappa|^4 |m_2|^4 E^2}{2M_1^2 M_2^4},
\eeq
predicting a power-law suppression proportional to $E^2$ at low energies. 

The expression (\ref{disp}) allows us to estimate the energy $E_{\rm trans}$ where the 
behavior of $\delta c_\nu$ changes from a quadratic growth to a constant. 
Equating
\be
\frac{|\delta|^4}{M_1^2} E^4\sim |\delta|^2 E^2\;,
\ee
we find 
\be
\label{Etrans}
E_{\rm trans}\sim M_1/|\delta|\;.
\ee
This shows that the transition occurs at energies parametrically larger
than $M_1$.

These analytic estimates are confirmed by a numerical calculation in which we compute exactly the dispersion relation $E_i=E_i(p)$ for the three neutrino states derived from
their Dirac equations. The neutrino group velocities is obtained by taking the derivative of $E_i$ with respect to momentum: $c_i = dE_i/dp$.
The numerical result corresponding to the choice $M_2=2M_1 =40\GeV$, $\theta_2 = 0.1$, $\kappa = 0.7$ is shown in \fig{mod}a. 
 
This model has a potential problem.
  The contribution to active neutrino masses in eq.~(\ref{LlowE})
  obtained by integrating out $N_1$ and $N_2$ is too large. Of course this can be cancelled by other contributions to Majorana masses of active neutrinos. Although this may seem to be a special fine tuning, the cancellation can occur naturally. One example is the case in which the active neutrinos have Yukawa couplings to a single sterile heavy neutrino $N_3$, while $N_{1,2}$ mix with $N_3$ in the mass matrix.  This generates a rank-one mass matrix for the lighter states. So this problem can be easily bypassed.

Another difficulty is that the model produces at low energy ($E \ll M_{1,2}$) an effect suppressed only by $E^2$, which is not enough to fit OPERA compatibly with the SN1987A bounds, as illustrated in \fig{mod}a. 
This problem can be solved in different ways, as shown by the following examples. 

\subsection{A model with quadratic mixing term}\label{mod2}
To remedy to the $E^2$ low-energy behavior of $\delta c_\nu$ we can add a further derivative in the Lorentz-violating term in eq.~(\ref{buf})
and use the Lagrangian 
\beq
\begin{array}{rl}
\Lag=& i\bar\n\bar\s^\m \d_\m\n+i\bar N_1\bar\s^\m\d_\m N_1+i\bar N_2\bar\s^\m\d_\m N_2
+\bigg(\displaystyle \frac{M_1}{2}N_1^2+\frac{M_2}{2}N_2^2+m_2 \n N_2+\mathrm{h.c.} \bigg)
\\
&+\kappa \, u^\n u^\m \d_\n N_1 \d_\m N_2
+\kappa^*\, u^\n u^\m \d_\n \bar N_1\d_\m \bar N_2\;.
\end{array}
\eeq

In the high-energy region ($E\gg M_2$) $\delta c_\nu$ is suppressed and dies out. In the intermediate region $M_1\ll E\ll M_2$ (or more precisely $E_{\rm trans}\ll E\ll M_2$, where 
$E_{\rm trans}$ will be computed in eq.~(\ref{Etrans2}) below), we obtain 
the effective Lagrangian
\beq
\label{LVE22}
\Lag_{\rm Lorentz}=\frac{\kappa m_2}{M_2} u^\n u^\m \;\n\,\d_\n\d_\m N_1
+\mathrm{h.c.}\;,
\eeq
where we have kept terms up to two derivatives.
This produces the dispersion relation 
\beq
\label{dispE2}
E^2-p^2-\frac{|\kappa|^2 |m_2|^2}{M_2^2}E^4=0\;.
\eeq
Hence
\beq
\delta c_\nu = \frac{3|\kappa|^2 |m_2|^2E^2}{2M_2^2}.
\eeq
We see that in this case the velocity of neutrinos is not constant in
the intermediate range but depends quadratically on energy.

At low energies ($E\ll M_1$) we integrate out the remaining sterile neutrino and
obtain
\beq
\label{LVE21}
\Lag_{\rm Lorentz}=-\frac{\kappa^2 m_2^2}{2M_1 M_2^2}(u^\m u^\n\d_\m\d_\n\n)^2\;.
\eeq
This yields the dispersion relation
\beq
\label{dispE4}
E^2-p^2-\frac{|\kappa|^4 |m_2|^4}{M_1^2 M_2^4}E^8=0
\eeq
and hence a $\delta c_\nu$ scaling as the sixth
power of energy,
\beq
\delta c_\nu = \frac{7|\kappa|^4 |m_2|^4E^6}{2M_1^2 M_2^4}\;.
\eeq
By comparing eqs.~(\ref{dispE4}) and (\ref{dispE2}) we obtain
that the transition between the two regimes occurs at 
\beq
\label{Etrans2}
E_{\rm trans}\sim \sqrt{\left| \frac{M_1 M_2}{\kappa m_2}\right|}\;.
\eeq

These analytic estimates are confirmed by our numerical calculation, as
exemplified in \fig{mod}b, which corresponds to the choice $M_2=1\TeV$, $M_1 =1\GeV$, $\theta_2 = 0.02$, $\kappa = (30\GeV)^{-1}$ (in the two-derivative case $\kappa$ is a dimensionful coupling). In this case we obtain a satisfactory fit of data and constraints, because of the steeper dependence on $E$ at low energies.

\subsection{A model with three sterile neutrinos}

An alternative approach is to employ  three species
of sterile neutrinos with Lagrangian 
\be
\label{3NL}
\Lag=i\bar\nu\bar\sigma^\m \d_\m\n+\bar N_i\bar\s^\m\d_\m N_i
+\bigg( m_i\n N_i+\frac{M_i}{2} N_i^2
+\sum_{i<j}i\kappa_{ij} u^\m N_i\d_\m N_j+\mathrm{h.c.}\bigg) \;,
\ee
where $i,j=1,2,3$. We assume the hierarchy of masses
$M_1\ll M_2\ll M_3$. 

To simplify the calculations let us
assume
$m_1=m_2=\kappa_{12}=0$. At large energies $E\gg M_3$ the active neutrino
decouples from the steriles and propagate with the speed of light. At 
$M_2\ll E\ll M_3$ we integrate out the third neutrino and obtain the
following Lorentz-breaking contributions
\be
\label{LB23}
\begin{array}{rl}
\Lag_{\rm Lorentz}=&\displaystyle \frac{\kappa_{13}m_3}{M_3}iu^\m\n\d_\m N_1
+\frac{\kappa_{23}m_3}{M_3}iu^\m\n\d_\m N_2\\
&\displaystyle+\frac{\kappa^2_{13}}{2M_3}(u^\m\d_\m N_1)^2
+\frac{\kappa^2_{23}}{2M_3}(u^\m\d_\m N_2)^2
+\frac{\kappa_{13}\kappa_{23}}{M_3}u^\m u^\n \d_\m N_1 \d_\n N_2\;.
\end{array}
\ee 
The velocity of the active neutrino in this regime is given by 
\be
\label{v23}
\delta c_\nu=|\delta_{13}|^2+|\delta_{23}|^2\;,~~~~~~
\delta_{13}\equiv \frac{\kappa_{13}m_3}{M_3}~,~~~
\delta_{23}\equiv \frac{\kappa_{23}m_3}{M_3}\;.
\ee

At lower energies
($M_1\ll E\ll M_2$)
we integrate out the second neutrino and the Lorentz breaking part becomes
\be
\label{LB12}
\Lag_{\rm Lorentz}=\frac{\kappa_{13}m_3}{M_3}iu^\m\n\d_\m N_1
+\frac{\kappa^2_{13}}{2M_3}(u^\m\d_\m N_1)^2
+\frac{\kappa_{23}^2m^2_3}{2M_2M_3^2}(u^\m\d_\m \n)^2\;,
\ee
where we have considered only terms up to two derivatives.
The third term here gives the quadratic growth of $\delta c_\nu$ with energy.

Finally, at $E\ll M_1$ we integrate out $N_1$ and obtain 
\be
\label{LB01}
\Lag_{LV}=\frac{m_3^2}{2M_3^2}\bigg[\frac{\kappa_{13}^2}{M_1}
+\frac{\kappa_{23}^2}{M_2}\bigg]
(u^\m\d_\m \n)^2\;.
\ee
The quadratic growth can be shut off at the price of the fine-tuning condition
$(\kappa_{13}/\kappa_{23})^2\simeq -M_1/M_2$.
The next operator is quartic in derivatives and will give an
$E^6$ dependence of $\delta c_\nu$ on energy. 

Let us conclude our discussion on these models with an important consideration.
 The models described here realize the optimal situation in which $\delta c_\nu$  is large only at OPERA energies: at lower energies there is a power-law suppression and
at higher energies $\delta c_\nu$ decreases because energy dominates over
the active/sterile mass mixing term.  In spite of that, the quantum corrections to the velocity of charged leptons $\ell$ is too large. For instance, in the model of sect.~\ref{mod1}, we find
\beq \delta c_\ell \approx \frac{\lambda^2_2}{(4\pi)^2}\kappa^2 \eeq
in agreement with the generic estimate of \eq{estimate}.
Note that supersymmetry cannot be invoked as a solution, because at momenta of the order of the OPERA energies the wino loop contribution cannot  cancel the $W^\pm$ loop.

\section{Conclusions}\label{concl}
It has been often speculated that Lorentz invariance can be violated in new frameworks beyond the Standard Model, especially in the context of various quantum gravity scenarios, such as space-time foam~\cite{ell}, string theory~\cite{kos}, Ho\v{r}ava gravity~\cite{hor}, extra dimensions with asymmetric space-time warping factors~\cite{gro}, or inhomogeneous Higgs backgrounds~\cite{grs}.
Tests of Lorentz invariance  can provide us with precious hints of physics at energy scales much beyond those directly probed in colliders and the experimental discovery of an actual Lorentz violation in nature would have a revolutionary impact on our understanding of the fundamental principles of the particle world.

The recent OPERA measurement, pointing towards evidence for superluminal propagation of neutrinos, opens an exciting perspective. Careful scrutiny of the OPERA methodology and confirmation from other experiments is certainly required before the measurement can be accepted. From the theoretical point of view, the result is quite surprising, not because of the violation of a sacred principle of special relativity, but because of the size of the effect. The amount of Lorentz violation in the neutrino sector required to explain the effect is much larger than the typical bounds that have been extracted for other Lorentz-violating parameters. This poses a question of logical consistency of the new experimental result.

In this paper, we have shown that the OPERA measurement is not necessarily inconsistent with other tests of special relativity in the neutrino sector, but requires very particular assumptions on $\delta c_\nu$, the relative difference between the neutrino and photon limit velocities. Various observational constraints, reviewed in sect.~\ref{sum}, force $\delta c_\nu$ to be nearly flavor universal, to drop steeply at low energies ($E_\nu < 10$~GeV) and remain roughly constant in the OPERA energy range (around 10 to 40 GeV). Such a peculiar behavior is unexpected but, as shown in sect.~\ref{mod} with the help of several illustrative models, can be accounted for by introducing sterile neutrinos with Lorentz-violating interactions and masses in the OPERA energy range.

As discussed in sect.~\ref{sum}, a fairly generic and formidable constraint on any interpretation of the OPERA measurement comes from $\delta c_e$, the relative difference between electron and photon limit velocities. Even if we restrict the original Lorentz violation to the neutrino sector, quantum corrections due to weak interactions will generally pollute the electron sector as well. Barring unknown cancellations, the resulting effect is too large, showing conflict between present bounds on $\delta c_e$ and the OPERA measurement. Fine-tuning the value of $\delta c_e$ at a particular energy cannot cure the problem, because there are strong bounds on $\delta c_e$ applicable at several different energies.

Much work has still to be done, both on the experimental and the theoretical front, before the fate of the OPERA measurement and its impact on physics will be determined.

\bigskip

{\bf Note added}

After this paper was published, the authors of ref.~\cite{cg} presented another strong constraint on superluminal interpretations of the OPERA result, based on energy loss from neutrino decay.
Their constraint is as numerically  stringent as the constraint presented here, based on loop contributions to $\delta c_e$,
but more direct.  Ref.~\cite{cg} assumed an energy-independent $\delta c$, but their argument
is fairly robust, since any dispersion relation leading to a superluminal neutrino,
like the ones we consider,  implies that a neutrino decay process  is kinematically allowed. Indeed, let us consider the decay process $\nu(p)  \to \nu(p')  X$, where $X=e^+e^-$ or any light final state.
The process is allowed when the invariant mass of $X$ is positive or, in other words, when the difference of the 4-momenta of the two neutrinos is time-like,
\beq
 E(p) - E(p^\prime )  > p-p^\prime .
\eeq
Here $E(p)$ is the neutrino energy for a generic dispersion relation and $p$ ($p^\prime$) is the momentum of the initial-state (final-state) neutrino. For a superluminal neutrino we approximately find
\beq
c > 1~~ \Rightarrow ~~c=\frac{dE}{dp} \approx \frac{ E(p) - E(p^\prime )}{p-p^\prime} >1.
\eeq
Thus a superluminal neutrino always has kinematically-allowed decay channels, independently of its dispersion relation.

\bigskip

{\bf Acknowledgements}

We thank Oriol Pujolas and Sergey Troitsky for useful discussions. S.S. is grateful to the CERN Theory Group for warm hospitality during his visit. The work of S.S. was supported in part by the Grants of the President of Russian Federation NS-5525.2010.2 and MK-3344.2011.2~(S.S.) and the RFBR grants 11-02-92108, 11-02-01528.    This work was supported by the ESF grant MTT8 and by SF0690030s09 project.

\end{document}